\documentclass[aps,  showpacs]{revtex4}

\begin{document}
\title{Gordon decomposition of Dirac current: a new interpretation}
\author{S. C. Tiwari \\
Department of Physics, Institute of Science,  Banaras Hindu University,  and Institute of Natural Philosophy, \\
Varanasi 221005, India }
\begin{abstract}
Historically Gordon decomposition of Dirac current played an important role in the interpretation of Dirac equation. We revisit it to understand the correspondence between Maxwell-Dirac and Maxwell-Lorentz theories. Arguments are presented to show that classical charge current corresponds to Gordon current. Consistency with Maxwell-Dirac theory leads to a new result: antisymmetric spin tensor in spin magnetization current of Dirac electron must be hidden as internal part of classical Maxwell tensor. It becomes natural to extend our previous interpretation in which electromagnetic field tensor represents angular momentum of photon or spacetime fluid (aether!) in the presence of sources such that the duality between field and source disappears. Both Gordon current and spin current are proposed to have 'spinning origin'. Formal expression of Gordon current makes it possible to apply the idea of optical vortex in this case, and quantized vortex and quantized orbital angular momentum  could be related with the conjecture of mechanical interpretation of electric charge. Separate mass and charge centers for spinning point charge are well known in the literature; we discuss them in the context of two-vortex model of electron and Gordon decomposition. 
\end{abstract}
\pacs{03.65.Pm, 12.60.Rc, 14.60.Cd}
\maketitle
\section{\bf Introduction}
Recent advances in spintronics and new developments in condensed matter systems have seen revival of interest in Dirac equation \cite{1}. On the other hand, in the context of modern quantum field theory (QFT) and the Standard Model (SM) of particle physics \cite{2} Dirac's relativistic wave equation in its original interpretation \cite{3} would seem rather obsolete and of merely historical interest. However quest for models beyond SM and need for new ideas for unification, in a sense, do recognize the foundational issues in QFT and drawbacks in SM. We have argued that revisiting the old questions afresh could throw light on some of the unresolved problems of today \cite{4}. Dirac equation had its origin in the strict adherence to the general principles of quantum mechanics and relativity, it would be interesting to see if there exists a  possibility to understand more about the nature of electron from this equation. Historically the enigma of spin and spin magnetic moment found nice physical interpretation in Dirac's theory; in the present paper we discuss new insights on the meaning of charge based on the Gordon decomposition of Dirac current \cite{5}.

Usually Gordon identities in the textbooks on QFT, see for example, \cite{6} are mentioned in connection with the calculations of electron vertex functions. However in the early days Gordon decomposition played a significant role in the physical interpretation of Dirac theory and its correspondence with nonrelativistic quantum theory \cite{7, 8, 9}. Contrary to the still prevailing belief that spin, being a typically nonclassical intrinsic property of a point electron, does not have a simple physical picture at least three notable contributions can be found \cite{10, 11, 12} that explain spin in terms of a circulating flow of electron wave field. Authors are careful to emphasize that they do not seek any model of electron with internal structure. Remarkably, amongst others, Barut draws deep insights into the internal structure of electron using zitterbewegung of free electron \cite{13, 14}. Unfortunately in all these works it has been assumed that electronic charge and electric current as understood in Maxwell-Lorentz theory need no further explanation: One simply interprets Dirac current as charge current. Departing from this conventional wisdom we argue that the source term in Maxwell equations in vacuum represents only a partial Dirac current, namely the so called Gordon current, and the spin tensor is hidden in the definition of the electromagnetic field tensor. Further assuming a finite-sized transverse wave field solution for Dirac wave function a vortex-like structure of electronic charge is envisaged.

In the next section preliminary aspects of Dirac relativistic wave equation and Gordon decomposition are summarized. The proposition that Gordon current corresponds to the classical charge current and classical electromagnetic field contains the spin tensor in hidden form is elucidated in Sec. III. Vortex solution in a simple picture is discussed in Sec.IV. Possible consequences and concluding remarks constitute the last section.
\section{\bf Preliminaries on Dirac relativistic equation}
Dirac's book \cite{2} continues to be a standard reference material on his relativistic equation for electron. Another important exposition of Dirac's theory could be found in \cite{7}. We may approach the derivation of Dirac equation using variational principle for the Lagrangian density
\begin{equation}
 L = \frac{i\hbar}{2} [\bar{\Psi} \gamma ^\mu \partial_\mu \Psi - (\partial_\mu \bar{\Psi}) \gamma^\mu \Psi] - m c \bar{\Psi} \Psi
\end{equation}
We have the notations: Lorentz 4-vectors $x^\mu = (x^0, {\bf {x}}),~ x_\mu =(x^0, -{\bf {x}}), ~ \partial_\mu = (\frac{\partial}{\partial x_0}, ~{\bf \nabla}), ~x_0 = ct, ~p^\mu p_\mu = E^2 -{\bf p}.{\bf p} = m^2 c^4$. Here the Greek indices are assumed to run over from 0 to 3. Dirac matrices are related by the definitions
\begin{equation}
 {\bf \alpha} = \gamma^0 {\bf \gamma}
\end{equation}
\begin{equation}
 \beta = \gamma^0
\end{equation}
and relativistic adjoint is
\begin{equation}
 \bar{\Psi} =\Psi^\dagger \gamma^0
\end{equation}
where Hermitian adjoint of $\Psi$ is $\Psi^\dagger$. Dirac equation for $\Psi$ and Dirac Hamiltonian are given by
\begin{equation}
 (i\hbar \gamma^\mu \partial_\mu - mc)\Psi = 0
\end{equation}\begin{equation}
 H = -i\hbar c~ {\bf \alpha}.{\bf \nabla} +m c^2 \beta
\end{equation}
Note that Eq.(5) is a Lorentz-covariant form of $H\Psi = E\Psi$ or
\begin{equation}
 (-i\hbar c~ {\bf \alpha}.{\bf \nabla} +m c^2 \beta)\Psi = i\hbar \frac{\partial \Psi}{\partial t}
\end{equation}

One can show that conserved Noether current corresponding to global gauge symmetry of (1) is Dirac current
\begin{equation}
 J^\mu_D = e \bar{\Psi} \gamma^\mu \Psi
\end{equation}
and for spacetime translational symmetry conserved canonical energy-momentum tensor is
\begin{equation}
 T^{\mu\nu} =\frac{\hbar c}{2}[\bar{\Psi} \gamma ^\mu \partial^\nu \Psi - (\partial^\mu \bar{\Psi}) \gamma^\nu \Psi]
\end{equation}
This tensor is not symmetric, however following Tetrode \cite{7} and also Belinfante \cite{15} a symmetric energy-momentum tensor could be constructed
\begin{equation}
 E^{\mu\nu} =\frac{1}{2} (T^{\mu\nu} + T^{\nu\mu})
\end{equation}
The symmetric tensor $E^{\mu\nu}$ also satisfies the usual conservation law \cite{16}.

Textbooks point out that $\bar{\Psi} \Psi$ in the Dirac current (8) can be interpreted as probability density and it is evidently positive definite. Most revealing is, however splitting Dirac current into two parts \cite{5}
\begin{equation}
 J^\mu_D = J^\mu_G +J^\mu_M
\end{equation}
\begin{equation}
 J^\mu_G = i \mu_B [\bar{\Psi} \partial^\mu \Psi - (\partial^\mu \bar{\Psi}) \Psi]
\end{equation}
\begin{equation}
 J^\mu_M =i \mu_B \partial_\nu M^{\mu\nu}
\end{equation}
Here the Bohr magneton is
\begin{equation}
 \mu_B = \frac{e\hbar}{2mc}
\end{equation}
and the anti-symmetric tensor
\begin{equation}
 M^{\mu\nu} = \frac{1}{2} \bar{\Psi}[\gamma^\mu \gamma^\nu  - \gamma^\nu \gamma^\mu] \Psi
\end{equation}
It could be easily verified that not only the Dirac current but the components (12) and (13) separately satisfy the continuity equations
\begin{equation}
 \partial_\mu J^\mu_G =0
\end{equation}
\begin{equation}
 \partial_\mu J^\mu_M =0
\end{equation}
Suffixes G and M indicate Gordon and spin magnetization current respectively. Physical interpretation of two currents is nicely discussed in \cite{7} and Sakurai's book \cite{8}; for a thought provoking novel discussion we refer to Gurtler and Hestenes \cite{11}.

In the next two sections an attempt is made to gain new insights on the electron model \cite{17, 18} based on the interpretation of electromagnetic field tensor as angular momentum tensor of photon fluid \cite{19} and the above decomposition (11) - (13).

\section{\bf Correspondence between Maxwell-Dirac and Maxwell-Lorentz theories}
Dynamics of charged bodies under the application of electromagnetic fields, free radiation fields, and radiation reaction on the motion of charged particles broadly constitute the electromagnetic phenomena. This picture of source and field together with their approximate separation and duality has served well the description of electromagnetic phenomena in classical electromagnetism. Essentially a similar paradigm is at work in quantum electrodynamics (QED): perturbative calculations and renormalization prescription to some extent provide a self-consistent treatment of self-fields and radiative corrections in the framework of quantized matter and radiation \cite{6}. However infinities, the meaning of external field and the nature of electron remain unsatisfactory at a fundamental level. Wheras the self-field and energy of a point electron turn out to be infinite in classical description, in addition to them electric charge also becomes infinite in QED. Origin of these problems could be traced to the source-field or matter-field duality; many failed attempts at unification testify to it. Could revisiting Dirac theory throw light on these iisues? In this section we address this question in detail.

${\bf A. ~Classical~ limit~ of~ currents~ in~ Dirac~ theory}$

Dirac current is a source for electromagnetic field, note however that the interpretation of Dirac current $J^\mu_D$ as electric charge current density would be vacuous unless one considers interaction with the electromagnetic field. In one of the standard methods one proceeds with the requirement of the local gauge invariance of the Lagrangian density: it becomes necessary to introduce a vector gauge field $A^\mu$ that couples to the Dirac field and transforms appropriately under the gauge transformation. Adding a kinetic energy term and assuming mass to be zero (for gauge invariance) the gauge field becomes a genuine dynamical field. Thus the total Lagrangian comprises of three parts
\begin{equation}
 L = L_D + L_M +L_{int}
\end{equation}
Defining the antisymmetric electromagnetic field tensor
\begin{equation}
 F^{\mu\nu} = \partial^\mu A^\nu - \partial^\nu A^\mu
\end{equation}
the Maxwell part of (18) is
\begin{equation}
 L_M = -\frac{1}{4} F^{\mu\nu} F_{\mu\nu}
\end{equation}
and interaction Lagrangian is
\begin{equation}
 L_{int} = -e \bar{\Psi} \gamma^\mu \Psi A_\mu
\end{equation}
Dirac equation in external field and inhomogeneous set of Maxwell equations follow from the variational principle as Euler-Lagrange equations of motion for the action
\begin{equation}
 S= \int L ~ d^4 x
\end{equation}
Explicitly, we have
\begin{equation}
 \gamma^\mu (i\hbar \partial_\mu - \frac{e}{c} A_\mu ) \Psi -m c \Psi =0
\end{equation}
\begin{equation}
\partial_\nu F^{\mu\nu} = e \bar{\Psi} \gamma^\mu \Psi
\end{equation}
In the discussions on physical interpretation of Dirac theory and its applications from 1928 onwards Eq. (23) has been of main focus \cite{3, 7, 8, 20, 21}.

Let us call the set of Eqs. (23) and (24) the Maxwell-Dirac theory. In classical electrodynamics the Maxwell-Lorentz theory is represented by
\begin{equation}
 \partial_\nu F^{\mu\nu}_c = J^\mu_c
\end{equation}
and the relativistic generalization of Newton-Lorentz equation of motion of electron
\begin{equation}
 m \frac{d v^\mu}{d \tau} = e v_\nu F^{\mu\nu}_c
\end{equation}
Here $\frac{d x^\mu}{d \tau}$ is velocity 4-vector and $d \tau$ is proper time. Suffix c is inserted to indicate 'classical'. It is crucial to recognize that the formal mathematical structure of the Maxwell-Lorentz theory is abstracted from experiments in which spin of electron played no role and current density $J^\mu_c$ represents charge flow of spinless electrically charged particle. Spin-dependent force is absent in Eq. (26). Thus the fundamental question raised here is the delineation of the correspondence between the set of Eqs. (23) and (24) and Eqs. (25) and (26). One of the remarkable characteristics of Maxwell equations is that these are inherently relativistically invariant. Obviously taking nonrelativistic limit of Dirac equation would serve no purpose. Other possibility is to find out quantum to classical reduction in the limit of Planck constant tending to zero.

Let us assume
\begin{equation}
 \Psi = C e^{\frac{iS}{\hbar}}
\end{equation}
and expand 4-component column vector $C$ in powers of Planck constant
\begin{equation}
 C= C_0 +\frac{\hbar}{i} C_1 + higher~ order ~powers ~in~ \hbar
\end{equation}
Note that phase in (27) $S$ has to be a scalar function and the amplitudes $C_i$ in (28) are column vectors. Substituting (27) along with the expansion (28) in the expressions (8), (12) and (13) for the currents and taking the limit $\hbar \rightarrow 0$ we arrive at the following
\begin{equation}
 J^\mu_{D, cl} = e ~\bar{C_0} \gamma^\mu C_0
\end{equation}
\begin{equation}
 J^\mu_{G, cl} =- \frac{e}{mc} \bar{C_0} C_0 \partial^\mu S
\end{equation}
\begin{equation}
 J^\mu_{M, cl} =0
\end{equation}
It has to be realized that Dirac current (8) and Gordon decomposition have been obtained for a free electron Dirac equation. In the presence of electromagnetic interaction we must use Eq. (23), however it turns out that the form of Dirac current (8) remains unchanged and 4-vector potential $A^\mu$ does not appear in the current. In this context it may be recalled that in both Schroedinger and Pauli theories the vector potential is present in the probability current. For example, we write the expressions for probability density and probability current density in Schroedinger theory below
\begin{equation}
 \rho_s = \Psi_s^* \Psi_s
\end{equation}
\begin{equation}
 {\bf J}_s = -\frac{i\hbar}{2m} (\Psi_s^* {\bf \nabla} \Psi_s -{\bf \nabla} \Psi_s^* \Psi_s) - \frac{e}{mc} {\bf A} \Psi_s^* \Psi_s
\end{equation}
Here $\Psi_s$ is Schroedinger wave function. Interestingly, performing Gordon decomposition in the presence of interaction while $J^\mu_M$ remains unaltered Gordon current (12) is modified to
\begin{equation}
 J^\mu_{G, int} = i\mu_B [\bar{\Psi} \partial^\mu \Psi - (\partial^\mu \bar{\Psi}) \Psi] -\frac{e^2}{mc^2} A^\mu \bar{\Psi} \Psi
\end{equation}
Formal similarity between ${\bf J}_{G, int}$ in (34) and Schroedinger current (33) is noticeable. Dropping $e$ in (34) could one interpret Gordon current as probability current? Unfortunately the time component of (34) would not be positive definite making this interpretation questionable, however in the nonrelativistic approximation such an interpretation has been suggested \cite{8}. Gurtler and Hestenes \cite{11} seeking consistency in Schroedinger, Pauli and Dirac theories argue that the widely accepted probabilistic interpretation of Dirac current also is inconsistent with Schroedinger theory; more startling is their conclusion that $\Psi_s$ in Schroedinger equation actually represents an electron in a spin eigenstate not a spinless particle.

It seems strange that in all these discussions on the interpretation \cite{2, 7, 8, 11} the consistency with Maxwell equation is not examined. For the Schroedinger current it is well known that substituting (33) for ${\bf J}_c$ in Macwell field equation (25) a phenomenological explanation of superconductivity can be given. We believe that limits of a theory, approximations to a theory and theory reduction are intricate but fruitful issues, see Ch. 9 in \cite{22}, nevertheless experiments and empirical facts ultimately have to be accorded the decisive role. Electron spin does have experimental evidence for its physical reality even if its concrete picture remains elusive, therefore we expect that experimental laws embodied in (25) and (26) hide spin effects in some way.

Since Dirac current and spin current remain unchanged in the presence of interaction their classical limits (29) and (31) would also not change. On the other hand the classical limit of Gordon current (34) making use of (27) and (28) assumes interesting form
\begin{equation}
 J^\mu_{G, cl, int} = -\frac{e}{mc} \rho (\partial^\mu S +\frac{e}{c} A^\mu)
\end{equation}
where we have defined
\begin{equation}
 \rho = \bar{C_0} C_0
\end{equation}
Note the difference between the classical limit of Dirac probability density $\bar{C_0} \gamma^0 C_0$ vide Eq. (29) and $\rho$ in (36). In the classical limit the probability interpretation is a nonissue and, therefore, one may identify (35) as the classical source term in (25). In that case a new question emerges: What is the significance of $A^\mu$ in the current density? 

${\bf B. ~Is~ Gordon ~ current~ classical~ current~?}$

If we take seriously the aforementioned interpretation we are led to ask: Does it indicate a sort of electromagnetic field structure of electron? It reminds us that beginning with the electron model of J. J. Thomson the idea that electron may have a purely electromagnetic origin, though unsuccessful, has never been abandoned. The form of (35) invites attention to Weyl's unified theory \cite{23}, Dirac's new electron theory \cite{24} and Weyl-Dirac theory \cite{25} generalized in \cite{26}. In Ch. 6 of electron monograph \cite{17} modifications of Maxwell equations are discussed in detail including the theories of \cite{23, 24}, however it comes out as a new realization that relativistic quantum equation of Dirac bears some resemblance with them. Let us elaborate.

In Weyl's theory \cite{23} Einstein and Maxwell field equations are derived in a unified way from the action functional proposed by Weyl; here we write the later one below
\begin{equation}
 \partial_\nu F^{\mu\nu} = constant~ A^\mu
\end{equation}
Since the current density is proportional to $A^\mu$ according to Weyl electric charge and current are 'diffused thinly throughout the world'. Amazingly in Dirac's 1951 theory \cite{24} in which gauge symmetry is broken imposing the nonlinear condition
\begin{equation}
 A_\mu A^\mu = k^2
\end{equation}
$k$ being a universal constant, Maxwell equation having Maxwell-Weyl form (37) is arrived at in which the constant is Lagrange multiplier, $\lambda$. Dirac's interpretation is that charge free case is $\lambda =0$ and for infinitesimal $\lambda$ neglecting changes in the fields the condition (38) replacing $A^\mu$ by $A^\mu+\partial^\mu S$ corresponds to the Hamilton-Jacobi equation for an electron if we set $k=m/e$.

Note that compared to Weyl theory \cite{23} and Dirac's new classical theory \cite{24} the Gordon current in classical limit (35) has a richer structure with the presence of additional functions $\rho$ and $S$. In this connection Maxwell equation derived in Weyl-Dirac theory postulating a new action function \cite{26} appears significant. It is well known that Weyl unified theory had been rejected on physical grounds, and almost went into oblivion in physics literature until Dirac revived it in 1973 \cite{25}. Besides Weyl's original monograph \cite{23} we refer to Dirac's article \cite{25} for a lucid account on Weyl geoetry; see also Ch. 8 in \cite{17}. Departing from the approach of Weyl and Dirac an in-invariant field (i. e. a scalar field of Weyl power zero) $\chi$ (in original article symbol $\Psi$ was used) was introduced to represent electron and the action was constructed from various curvatures and $\xi$
\begin{equation}
 \xi = \partial_\mu \chi \partial^\mu \chi
\end{equation}
Of present interest is Maxwell field equation in this generalized Weyl-Dirac theory that we reproduce directly
\begin{equation}
 \partial_\nu F^{\mu\nu} = constant ~ (\partial^\mu \xi +2 \xi A^\mu)
\end{equation}
Formal similarity of current density in (40) with (35) can be easily noticed; in a more suggestive form the right hand side of (40) can be rewritten as $\xi (\partial^\mu(ln \xi)/2 + A^\mu)$. It is straightforward to show that for $\xi = constant$, Eq. (40) reduces to Maxwell-Weyl form (37).

Differing physical motivations and underlying foundations in \cite{23, 24, 26} and the limiting Maxwell-Dirac theory with source current (35) seem to converge on a formal system in which 4-vector potential assumes fundamental significance representing even the charge current. Unfortunately in spite of vast efforts of many physicists this attractive approach to dissolve source and field separation has not succeeded. However, it may be noted that the preceding discussion shows that the correspondence between Maxwell-Dirac theory and Maxwell-Lorentz theory entails identification of Gordon current with classical charge current. In that case, how do we reconcile it with the standard interpretation of Eq. (24) as Maxwell equation with Dirac current as a source current? 

${\bf C.~ New ~ interpretation}$

A radical alternative presents itself if we shift focus of our attention to the possibility of modifying electromagnetic field tensor (19). Clues to proceed in this direction stem from mathematical arguments and physical insights obtained from different angles.

Given a 4-vector field $V^\mu$ satisfying the first order equation
\begin{equation}
 \partial_\mu V^\mu = 0
\end{equation}
one can express it in terms of an antisymmetric tensor field $G^{\mu\nu}$
\begin{equation}
 V^\mu = \partial_\nu G^{\mu\nu}
\end{equation}
Tensor $G^{\mu\nu}$ is not uniquely determined; one could add another antisymmetric tensor $H^{\mu\nu} = \partial^\mu U^\nu - \partial^\nu U^\mu$ to $G^{\mu\nu}$ in which 4-vector $U^\mu$ is required to obey the wave equation
\begin{equation}
 \partial_\nu \partial^\nu U^\mu = 0
\end{equation}
Implication of above simple mathematics in the present context becomes immediately evident recognizing that Dirac current (8), Gordon current, spin current and classical current all satisfy continuity equations of the form (41), and covariant Lorentz gauge condition too has similar form
\begin{equation}
 \partial_\mu A^\mu = 0
\end{equation}
In the standard theory, electromagnetic field tensor has physical reality in the sense that measurable quantities like energy, momentum and force involve the fields ${\bf E}$ and ${\bf B}$ explicitly and due to gauge freedom Eq. (44) has just a subsidiary role. It is only in quantum theory that physical reality of electromagnetic potentials manifests in typical phenomenon like Aharonov-Bohm effect. We have proposed that Eq. (44) is fundamental to describe photon physics \cite{27}. In such a scenario relavance of Eqs. (41) and (42) extends to both charge currents and fields.

Further scrutiny of Eq. (40) shows that for $\xi =0$, i. e.
\begin{equation}
 \partial_\mu \chi \partial^\mu \chi = 0
\end{equation}
fields $F^{\mu\nu}$ and $\chi$ decouple completely: in Weyl geometry $F^{\mu\nu}$ is called distance curvature by Weyl \cite{23} and in Eq. (45) $\partial^\mu \chi$ is interpreted null propagation vector \cite{17, 26}. Purely geometrical quantities need not have physical realization, however following Weyl and Dirac it is illuminating to relate Eqs. (40) and (45) with electromagnetic phenomena. Identification of distance curvature with electromagnetic field tensor in Weyl theory is well known; however the proposition that null vector describes electron would seem to contradict the established experiments. Closer examination reveals that this apprehension is unfounded as the experiments correspond to electron + electromagnetic fields either bounded or external radiation fields whereas in the generalized Weyl-Dirac theory \cite{26} electron is decoupled and truly free. 

In fact, this idea finds support in Dirac's theory in which it is well known that the velocity eigenvalues of electron are $\pm c$. Dirac's explanation invoking Heisenberg uncertainty principle and Schroedinger's zitterbewegung \cite{3} have been widely discussed in the literature the idea being to ascribe only abstract mathematical reality to this and assert that physical electron in Dirac theory has usual observed properties. Interestingly Pauli \cite{7} mentions that Breit was the first to point out that velocity and momentum of Dirac electron are not related as one would expect in classical theory. Corben's monograph \cite{28}, undoubtedly a thorough treatment of spinning point charged particle, emphasizes an important result: center of mass and center of charge may lie at different points and momentum need not be proportional to velocity. According to him due to 'the decoupling of the velocity from the momentum' a classical analogue of zitterbewegung could be envisaged such that the momentum is constant and velocity oscillatory. It is rather disconcerting to find that in spite of the emergence of a novel property, i. e. free electron velocity equal to the velocity of light, and a fundamental length scale, i. e. the Compton wavelength $\lambda_c =\hbar /mc$ for a Dirac electron, logical arguments are overstretched to deny them physical reality. In a refreshing contrast some physicists did try to draw insights from them \cite{10, 12, 13, 14}. Barut adopts unconventional approach and assuming internal structure of electron investigates observable effects of zitterbewegung: a fundamental length (finally identified with $\lambda_c$), spin as orbital angular momentum and rest mass as internal energy constitute principal results in \cite{13}, while in \cite{14} a classical model of spin is discussed. Barut and Zanghi \cite{14} stress a striking correspondence between Abraham-Lorentz-Dirac classical equation for spinless electron incorporating radiation reaction term with quantum Dirac equation and covariant symplectic classical system proposed in their paper. Barut and Bracken \cite{13} state unambiguously that Dirac equation admits substructure of electron, and suggest that possibly electron is not a point particle but 'a massless charge performing a complicated motion around a center of mass'. The nature of charge itself is not analyzed and the standard interpretation of Maxwell field is accepted by them; it could be that this prevented them from making further progress in the ingenious electron model of Barut.

Plausible arguments have been put forward above to identify the classical limit of Gordon current (35) with the classical source current in the Maxwell  equation. Now we propose that the complete expression of Gordon current corresponds to the experimentally determined electric charge current on the right of Maxwell equation (25). If we insist on the consistency with Maxwell-Dirac theory then an interesting interpretation could be given: spin magnetization tensor (15) is an internal hidden part of $F^{\mu\nu}_c$. Thus $F^{\mu\nu}_c$ differs from (19). Absorption of $M^{\mu\nu}$ in $F^{\mu\nu}_c$ should not be confused with the introduction of constitutive relations for electromagnetic field in material media.
Cogent arguments presented in Sec. 4 of \cite{19} led us to interpret Maxwell tensor as angular momentum tensor of what was termed as photon fluid. Dimensional analysis suggests that $\frac{\hbar}{e} F^{\mu\nu}_c$ and dropping $\frac{e}{mc}$ in (13) the tensor $\frac{\hbar}{2} M^{\mu\nu}$ acquires the dimension of angular momentum, therefore spin tensor as internal part of Maxwell tensor becomes quite natural.

Maxwell equations have interesting features and symmetries, the one that we highlighted in \cite{17} and elaborated in \cite{19} is largely unnoticed in the literature: the electronic charge $e$ can be factored out and becomes redundant. Geometric framework, for example, Weyl geometry has immediate application due to this. Not only this, wherever matter-radiation interaction is involved, e. g. Newton-Lorentz equation (26) or Dirac equation (23) factoring out $e$ from electromagnetic quantities one always ends up with the combination $\frac{e^2}{c}$ that curiously has the dimension of angular momentum. Attaching deep physical significance to it a conjecture was put forward that electric charge is a manifestation of mechanical rotation in \cite{17}. The present interpretation of $F^{\mu\nu}_c$ would be in harmony if just like $J^\mu_M$ the Gordon current is assigned 'spinning origin'.

To conclude: in the light of the proposed interpretation Maxwell field equation represents angular momentum continuity equation such that the distinction between charge-current and field disappears at a fundamental level.

\section{\bf Quantized vortex and Gordon current}
Mathematical formalism of particles and fields in one of the important forms proceeds with the second order linear wave equation; the simplest and the most common solution is that of idealized plane wave. A great deal of physical understanding can be obtained in the plane wave assumption. Though Dirac equation is a first order wave equation each component of Dirac spinor satisfies the second order wave equation. In the study of Dirac equation plane wave representation provides useful interpretational insights \cite{3, 7, 8}. If the need be, one could construct a wavepacket using Fourier decomposition and momentum space expansion. Assuming Gaussian wavepacket \cite{10, 12} for Dirac electron a picture of circulating wave orbital angular momentum for spin has been given. We draw attention to the fact that for past few decades propagating field singularities as solutions of second order wave equations have been of great interest in optics. In analogy with crystal defects a phase singularity where amplitude of a complex scalar field is zero and phase indeterminate is termed a dislocation. In the fluid dynamical analogy such a propagating singularity or topological defect is called a vortex. One of the most studied and fruitful phase singularities in optics is that of Laguerre-Gaussian (LG) beams \cite{29} which have finite transverse size and helicoidal wavefront, azimuthal phase $e^{il\phi}$ and propagation along z-axis $e^{ikz - i\omega t}$. Here one works in cylindrical coordinate system $(r, \phi, z)$ and the integer $l$ is a topological index. For vector waves the notion of disclinations has been introduced \cite{30}. These are termed vector vortices in fluid analogy: topological photon proposed recently \cite{27} is a propagating vector vortex. The purpose of this section is to outline a vortex-like rendition of Gordon current.

The main idea is an extension of the logic developed in \cite{31}: the form of time-averaged Poynting vector calculated for paraxial wave in \cite{29}
\begin{equation}
 {\bf S} = \frac{i\omega \epsilon_0}{2} [u {\bf \nabla} u^* -u^* {\bf \nabla} u -2 i k u^* u \hat{z}]
\end{equation}
transformed to a covariant form resembles the Noether current for U(1) gauge symmetry for a massless complex scalar field $\Phi$
\begin{equation}
 C^\mu = i [\Phi^* \partial^\mu \Phi - \Phi \partial^\mu \Phi^*]
\end{equation}
To derive (46) assuming paraxial wave equation for the vector potential ${\bf A} = \hat{x} u(x, y, z) e^{ikz - i\omega t}$ the electric and magnetic fields are calculated from the defining relations and finally the Poynting vector ${\bf S} = \epsilon_0 {\bf E} \times {\bf B}$ is calculated giving expression (46).

Gordon current (12) has a look-a-like form of (47) and the components of $\Psi$ also satisfy the second order wave equation besides Dirac equation (5); tentatively a vortex structure in analogy with optical vortex could be envisaged. Complications would stem from $\Psi$ being a spinor, however to illustrate the proposed idea we make a simplifying assumption that spinor and spacetime variables are separable. This assumption is akin to a plane wave solution of free Dirac particle $\Psi_i (x,t) = w_i e^{ik_\mu x^\mu}$ where $w_i,~ i=1, 2, 3, 4$ are numbers. Thus $\Psi$ is assumed to be a product of a constant spinor $D$ and a scalar function $\Phi_G$
\begin{equation}
 \Phi_G = u_0 (r, z)~ e^{il\phi} ~e^{ikz-i\omega t}
\end{equation}
Here amplitude $u_0 (r, z)$ comprises of associated Laguerre polynomial and Gaussian function in the case of LG modes. For the first doughnut mode $(l=\pm 1)$ the amplitude vanishes on the axis, i. e. at $r=0$ and the phase changes by $\pm 2\pi$ for any complete circuit around the axis. Gordon current and momentum density would be nonzero, and z-component of orbital angular momentum would turn out to be quantized in terms of the azimuthal index or topological charge $l$ \cite{29, 31}. Note that for a Gaussian wavepacket \cite{10, 12} Gordon current and momentum are zero. Spin finds interpretation as orbital angular momentum using spin current in \cite{10}, it is quite illuminating that another attribute of electron, namely electric charge could be interpreted in terms of orbital angular momentum using Gordon current. Interestingly this places two parts of Gordon decomposition of Dirac current (11) on equal footing (to avoid confusion we remark that momentum density differs from Gordon current by $\gamma^0$).

\section{\bf Discussion and Conclusion}

Hydrodynamics played a very significant role in the development of Maxwell theory, and fluid model of aether to represent physical reality had a distinguished contribution in the history of physics \cite{32}. We believe that in the modern era fluid dynamical analogy has inspiring value, and perhaps immense potential to offer an alternative to unification schemes based on gauge theories and superstrings \cite{2, 4}. In the beginning, extending particle interpretation of Schroedinger theory to Dirac relativistic equation, the so called single-particle interpretation though led to a profound prediction that of anti-particle (positron), it is now generally believed that second quantized (or q-number field theory) is imperative for acceptable physical interpretation. It has to be realized that Dirac equation could also be viewed as a c-number field theory. The fluid approach adopted in the present paper would be more close to classical Dirac field theory: the particles, essentially electron and photon here, are to be viewed as topological defects or quantized vortices. Interaction mechanism would correspond to vortex-vortex interaction and disturbances in the rotating spacetime fluid.

A definitive work in this approach has not yet been accomplished, however the advances made in \cite{17, 19, 27} give us confidence that learning from the past efforts of great thinkers mentioned in \cite{32} and current developments in field theories a viable mathematical formalism and physical picture would emerge. The present paper is focused on a very specific aspect of Dirac theory. Two ideas, in the preliminary stage of study, could be briefly discussed. The existence of separate charge and mass centers for a free Dirac electron, though intriguing, has been discussed in the literature \cite{13, 28}. Recalling that two fundamental lengths $\hbar/mc$ and $e^2/mc^2$ are associated with electron, in the light of two-vortex model \cite{18} the question whether the spin current and the Gordon current could correspond to a pair of vortices with the respective core dimensions seems attractive.

Another idea is concerned with the formalism. Field description of topological photon \cite{27} is based on the postulate that the Lorentz gauge condition (44) is the field equation for a photon, and Nye's disclination \cite{30} is applied to the vector potential. For a vector field propagating in a definite direction disclination is the locus of points where the transverse components of the vector field are zero. There are two equations (16) and (17) for electron similar to only one equation (44) for photon, however there is another crucial difference that vectors $J^\mu_G$ and $J^\mu_M$ have Dirac spinor as a basic field. How to generalize the concept of disclination in this case? It is a nontrivial problem. An interesting approach could be to exploit the fact that in the nonrelativistic limit 2-component spinor is adequate for both positive and negative energy solutions retaining large components. Writing $\Psi$ in terms of 2-component spinors $(\Psi_u, \Psi_l)$ and assuming propagation along z-axis a reasonable definition for spinor vortex could be the locus of points where
\begin{equation}
 \Psi_u =0 ~~ or~~ \Psi_l =0
\end{equation}
Thus we are replacing the vanishing of the transverse components of the vector field defining disclination by Eq. (49) for Dirac spinor. Could this approach help resolve negative energy problem in Dirac equation? Could one relate the sign of topological charge i. e. the sign of electric charge with the sign of energy? These questions need further investigation, and are receiving our attention.

In the present paper we have used some of the ideas from the photon model proposed in \cite{27}, and since recently a review of this paper \cite{33} has come to our attention it is desirable to make few remarks on the criticisms. Main criticism is regarding the motivation for a radical approach, explanation of experimental facts in new theory and verifiable predictions. Let it be reminded that precision experiments support the Standard Model of particle physics and yet we have been witnessing immense efforts for going beyond SM for past few decades \cite{2}: conceptual and foundational problems are prime motivating factors. In QED and Maxwell-Lorentz theory also there do exist such problems \cite{4, 17} and the concept of photon continues to remain enigmatic \cite{19}. Recent excellent reviews \cite{34} highlight controversies regarding momentum and angular momentum in QED and Quantum Chromodynamics. Theory reduction has important role: if an established theory could be shown as a limiting theory of the new one the empirical facts would not be contradicted in the new theory. A typical example is Einstein gravity and Newton gravity; however an entirely new picture may emerge in a new theory - pseudo-Riemannian spacetime geometry in Einstein theory. One may have merely a new interpretation without changing the formalism, for example scalar field of Schroedinger interpreted as probability amplitude in Born's interpretation of quantum theory. For more detailed discussion see \cite{22}. Previous discussions in \cite{19, 35} and the final section of \cite{27} show that our approach is not in conflict with the established experimental facts, provides clear physical picture for so called counter-intuitive and mysterious phenomena reported in the literature and has new predictions.

In conclusion, revisiting physical interpretation of Gordon decomposition of Dirac electron a new perspective on the correspondence between Maxwell-Dirac theory and Maxwell-Lorentz theory is developed, and the implications on the two-vortex model of electron are discussed.

\end{document}